\documentclass[showpacs,twocolumn,prl,aps]{revtex4}
\usepackage{graphicx}

\begin{document}

\title[Short title for running header]{Variational study of the neutron resonance mode in the cuprate superconductors}
\author{Tao Li$^{1}$, Fan Yang$^{2}$}
\affiliation{$^{1}$ Department of Physics, Renmin University of
China, Beijing 100872, P.R.China\\
 $^{2}$ Department of Physics,
Beijing Institute of Technology, Beijing 100081, P.R.China}
\date{\today}

\begin{abstract}
A Gutzwiller-type variational wave function is proposed for the
neutron resonance mode in the cuprate superconductors. An efficient
re-weighting technique is devised to perform variational Monte Carlo
simulation on the proposed wave function which is composed of
linearly superposed Gutzwiller projected Slater determinants. The
calculation, which involves no free parameter, predicts
qualitatively correct behavior for both the energy and the spectral
weight of the resonance mode as functions of doping.
\end{abstract}
\maketitle

\section{Introduction}
The $(\pi,\pi)$ resonance mode observed by neutron scattering in the
cuparte superconductors is among the most prominent phenomena in
these systems\cite{Fong,Fong1,Dai,He}. Below the superconducting
transition temperature $T_{c}$, a sharp peak is observed in the spin
fluctuation spectrum around $(\pi,\pi)$. The mode is found to have
close correlation with superconductivity of the system. For example,
the mode energy, which is temperature independent, is found to scale
linearly with $T_{c}$ as a function of doping. At the same time, the
intensity of the mode is found to have the similar temperature
dependence as the superfluid density.

Much theoretical efforts have been devoted to the understanding of
the origin of the neutron resonance mode and its correlation with
superconductivity\cite{Liu,Mazin,Norman,Lee,Weng,ZhangSC}. Among
these theories, the RPA-like theory, which takes the resonance mode
as a spin-one bound state in the particle-hole channel(spin exciton)
induced by the residual attractive interaction between the Bogliubov
quasiparticles in the superconducting state, is the most popular. In
this theory, the dynamical spin susceptibility is given by
\begin{equation}
\chi(\mathrm{q},\omega)=\frac{\chi_{0}(\mathrm{q},\omega)}{1-U(\mathrm{q})\chi_{0}(\mathrm{q},\omega)},
\end{equation}
here $\chi_{0}(\mathrm{q},\omega)$ is the bare spin susceptibility
of the BCS superconducting state determined by both the band
dispersion and the gap function. $U(\mathrm{q})$ is the
phenomenological RPA correction factor chosen to fit the
experimental data. In the RPA theory, the superconducting gap sets a
natural energy scale for the resonance mode below which the mode is
stable. At the same time, when the system approaches the
antiferromagnetic ordering instability under RPA correction, the
mode will evolve into the Goldstone mode of the ordered state and
its energy will approach zero. Thus, the energy of the resonance
mode in the RPA theory is determined by both the magnitude of the
superconducting gap and the strength of the antiferromagnetic
correlation.

Although the RPA theory can account for some aspects of the neutron
resonance mode, it is at the best, a phenomenological theory. The
band dispersion of the quasiparticle and the RPA correction factor
$U(\mathrm{q})$, on which the result of RPA calculation depends
sensitively on, are subjected to fine tuning. An effort to combine
the more microscopic slave Boson mean field theory and the RPA
theory in the $t-J$ model\cite{Lee} has resulted in too large a
doping range($0<x<0.2$, $x$ is doping concentration) in which the
system is unstable with respect to antiferromagnetic ordering. At
the same time, neither the phenomenological RPA theory nor the RPA
correction on the slave Boson mean field theory of the $t-J$ model
respects the local spin sum rule of the $t-J$ model, $\int dq
d\omega S(q,\omega)=(1-x)\frac{3}{4}$, as a result of their neglect
of the no double occupancy constraint of the $t-J$ model.

In this paper, we propose a variational description for the neutron
resonance mode with a Gutzwiller projected wave function. Our
approach can be taken as the generalization of the usual RPA theory
into the Hilbert space satisfying the no double occupancy constraint
of the $t-J$ model. The approach has the virtual that it is
parameter free: the RPA correction is automatically done through the
variational procedure.  We also devise an efficient algorithm to do
Monte Carlo simulation on the variational wave function we proposed.
Numerical calculation shows that our variational description of the
neutron resonance mode capture its basic characteristics very well.

The paper is organized as follows. In the next section, we introduce
the variational ground state and present the result of the single
mode approximation on which the result of our variational
calculation would compare to. We then introduce our variational wave
function for the neutron resonance mode and the numerical technique
to do Monte Carlo simulation on it. Then we present our numerical
results and offer a discussion on the results. Finally, we conclude
this paper with some further problems to be addressed in the future.

\section{The single mode approximation}
We take the $t-J$ model as the basic model to describe the physics
of high-$T_{c}$ superconductors
\begin{eqnarray}
H&=&H_{t}+H_{t'}+H_{J},\nonumber\\
H_{t}&=&-t\sum_{<i,j>,\sigma}(\hat{c}_{i,\sigma}^{\dagger}\hat{c}_{j,\sigma}+h.c.),\nonumber
\\
H_{t'}&=&-t'\sum_{<<i,j>>,\sigma}(\hat{c}_{i,\sigma}^{\dagger}\hat{c}_{j,\sigma}+h.c.),\nonumber
\\
H_{J}&=&J\sum_{<i,j>}(\vec{\mathrm{S}}_{i}\cdot
\vec{\mathrm{S}}_{j}-\frac{1}{4}n_{i}n_{j}).
\end{eqnarray}
Here $\hat{c}_{j,\sigma}$ is the constrained electron operator
satisfying the constraint
$\sum_{\sigma}\hat{c}_{i,\sigma}^{\dagger}\hat{c}_{i,\sigma}\leq 1$.
$\sum_{<i,j>}$ and $\sum_{<<i,j>>}$ represent the sum over nearest
neighboring(NN) and next nearest neighboring(NNN) sites. Here we
take $\frac{t'}{t}=-0.25$ to describe hole doped system. The
exchange term is fixed at $\frac{J}{t}=\frac{1}{3}$.

The no double occupancy constraint
$\sum_{\sigma}\hat{c}_{i,\sigma}^{\dagger}\hat{c}_{i,\sigma}\leq 1$
is crucial for the spin dynamics of the system. With this
constraint, the electron behave like mobile $s=\frac{1}{2}$ spin
rather than usual free electron. More quantitatively, the spin
structure factor in the $t-J$ model satisfies the following local
spin sum rule
\begin{eqnarray}
\int d\mathrm{q} d\omega S(\mathrm{q},\omega)=(1-x)\frac{3}{4},
\end{eqnarray}
in any physical state, here $x$ is hole density. When the constraint
is relaxed, as is done in slave Boson mean field theory or
phenomenological RPA theory, the spin fluctuation would be strongly
suppressed and no such sum rule would apply.

To satisfy the local spin sum rule of the $t-J$ model, the
variational ground state, on which to construct the variational
excitations, must respect the no double occupancy constraint. The
Gutzwiller projected d-wave BCS
state\cite{ZhangFC,Gros,Gros1,Ogata}, which satisfy the no double
occupancy constraint and for long has been known as an excellent
variational description of the ground state of the system, is the
most natural choice for this purpose.

Thus our variational ground state is given by
\begin{eqnarray}
| \Psi_{g} \rangle&=& P_{N} P_{G}|d-\mathrm{BCS} \rangle\nonumber
\\
&=&P_{G}(\sum_{i,j}a(i-j)c^{\dagger}_{i,\uparrow}c^{\dagger}_{j,\downarrow})^{\frac{N}{2}}|0\rangle
\end{eqnarray}
in which $P_{N}$ is the projection operator into the subspace with
$N$ electrons and $P_{G}$ is the Gutzwiller projection operator into
the subspace of no double occupancy,
$a(i-j)=\sum_{\mathrm{k}}\frac{v_{\mathrm{k}}}{u_{\mathrm{k}}}e^{i\mathrm{k}(\mathrm{r}_{i}-\mathrm{r}_{j})}$
is the real space wave function of the Cooper pair with
$\frac{v_{\mathrm{k}}}{u_{\mathrm{k}}}=\frac{\Delta_{\mathrm{k}}}{\xi_{\mathrm{k}}+E_{\mathrm{k}}}$.
Here,
$E_{\mathrm{k}}=\sqrt{\xi_{\mathrm{k}}^{2}+\Delta_{\mathrm{k}}^{2}}$
, $\xi_{\mathrm{k}}$ and $\Delta_{\mathrm{k}}$ are given by
\begin{eqnarray}
\xi_{\mathrm{k}}&=&-2(\cos \mathrm{k}_{x}+\cos
\mathrm{k}_{y})-4t'_{v}\cos \mathrm{k}_{x}\cos
\mathrm{k}_{y}-\mu_{v}\nonumber\\
\Delta_{\mathrm{k}}&=&2\Delta_{v}(\cos \mathrm{k}_{x}-\cos
\mathrm{k}_{y}),
\end{eqnarray}
in which $t'_{v}, \mu_{v}, \Delta_{v}$ are variational parameters to
be determined by the optimization of the ground state energy with
respect to the $t-J$ model. We note $t'_{v}, \mu_{v}, \Delta_{v}$
are just variational parameters, rather than real NNN hoping term,
real chemical potential and real superconducting gap.

Now we construct the spin excitation on the variational ground
state.  As a first approximation to $(\pi,\pi)$ resonance mode, we
adopt the single mode approximation of the form
\begin{eqnarray}
|\Psi_{Q}^{0}\rangle=\mathrm{S}_{Q}^{+}|\Psi_{g}\rangle,
\end{eqnarray}
in which $\mathrm{S}_{Q}^{+}$ is the creation operator of the spin
density excitation at $Q=(\pi,\pi)$. As the variational ground state
satisfy the no double occupancy constraint, the spin excitation
spectrum is guaranteed to obey the local spin sum rule.

The single mode approximation is a good approximation when the spin
fluctuation spectrum is dominated by the contribution form the
resonance mode. The excitation energy in the single mode
approximation can be calculated in the standard way
\begin{eqnarray}
E_{Q}=\frac{\langle\Psi_{Q}^{0}|H|\Psi_{Q}^{0}\rangle}{\langle\Psi_{Q}^{0}|\Psi_{Q}^{0}\rangle}-E_{g},
\end{eqnarray}
where $E_{g}$ denotes the ground state energy. Assuming that
$H|\Psi_{g}\rangle=E_{g}|\Psi_{g}\rangle$, we have
\begin{eqnarray}
E_{Q}=\frac{1}{2}\frac{\langle\Psi_{g}|[S_{Q}^{-},[H,S_{Q}^{+}]]|\Psi_{g}\rangle}{\langle\Psi_{g}|S_{Q}^{-}S_{Q}^{+}|\Psi_{g}\rangle}.
\end{eqnarray}
Using the commutation relation
\begin{eqnarray}
[\hat{c}_{i,\sigma}^{\dagger},S_{j}^{+}]&=&-\delta_{ij}\delta_{\sigma,\downarrow}\hat{c}_{i,\uparrow}^{\dagger}\nonumber
\\\
[\hat{c}_{i,\sigma},S_{j}^{+}]&=&\delta_{ij}\delta_{\sigma,\uparrow}\hat{c}_{i,\downarrow},
\end{eqnarray}
in which $\hat{c}_{i,\sigma}^{\dagger}$ is the constrained electron
creation operator at site $i$ and $S_{j}^{+}$ is spin lifting
operator at site $j$, the mode energy in the single mode
approximation can be shown to be given by
\begin{eqnarray}
E_{Q}=\frac{-\langle\Psi_{g}| H_{t}|\Psi_{g}
\rangle-\frac{8}{3}\langle\Psi_{g}|
H_{J}|\Psi_{g}\rangle}{\langle\Psi_{g}|S_{Q}^{-}S_{Q}^{+}|\Psi_{g}\rangle}.
\end{eqnarray}
It is interesting to note that the $H_{t'}$ term of the Hamiltonian
does not contribute when $Q=(\pi,\pi)$.

The mode energy calculated from Eq.(10) is shown in Fig. 1. Two
things are to be noted here. First, $E_{Q}$ increases monotonically
with doping, consistent with observation in the underdoped regime.
Second, $E_{Q}$ approaches zero at half filling in the thermodynamic
limit. The monotonic increase of $E_{Q}$ is due to the rapid
decrease of the spin structure factor as a function of doping which
overcompensates the increase of the absolute value of the kinetic
energy and the exchange energy. The second is in fact a reflection
of the Goldstone theorem. At half filling, the spin structure factor
calculated from $|\Psi_{g}\rangle$ scales superlinearly with the
number of lattice site $N_{s}$
($M_{Q}^{2}=\langle\Psi_{g}|S_{Q}^{-}S_{Q}^{+}|\Psi_{g}\rangle/\langle\Psi_{g}|\Psi_{g}\rangle\propto
N_{s}^{\frac{4}{3}}$) as a result of the long range correlation
induced by the Gutzwiller projection\cite{Gros1}. However, the
kinetic and exchange energy is by definition proportional to
$N_{s}$. Thus $E_{Q}$ must approach zero at half filling as
$N_{s}\rightarrow \infty$. This is to be compared with the mean
field prediction that $M_{Q}^{2}\propto N_{s}$ at all doping which
imply that $E_{Q}$ remains finite even at half filling. Thus the
Gutzwiller projection in our wave function plays a key role to
recover the correct trend of the mode energy as a function of
doping.

The mode energy calculated by the single mode approximation is in
fact the center of gravity of the spin fluctuation spectrum at
$Q=(\pi,\pi)$ and thus overestimates the energy of the resonance
mode which lies at the bottom of the spin fluctuation spectrum. In
fact, the single mode approximation we adopted above has nothing to
say about the very existence of the resonance mode. This is
especially clear at large doping when the mode energy predicted by
Eq.(10) stretches into the particle-hole continuum. Thus, although
the single mode approximation gives correctly the trend of the mode
energy as a function of doping, it is too crude to give a
quantitative answer on both the mode energy and mode weight.

It should be noted that the energy calculated from Eq.(10) with the
variational ground state is only a approximation to the single mode
approximation energy(as we have made the assumption that
$H|\Psi_{g}\rangle=E_{g}|\Psi_{g}\rangle$). Thus, although Eq.(10)
is by definition positive definite, the true single mode
approximation energy can be negative when the system become unstable
with respect to magnetic ordering at $Q=(\pi,\pi)$. We will
encounter this situation below when we calculate the single mode
approximation energy directly from Eq.(7).

\begin{figure}[h!]
\includegraphics[width=8cm,angle=0]{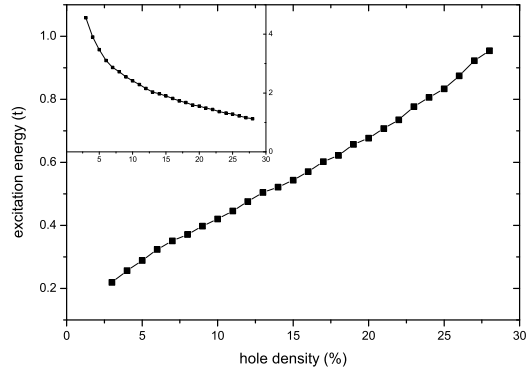}
\caption{The resonance energy determined from the single mode
approximation as a function of doping. The inset shows the structure
factor at $Q=(\pi,\pi)$ (divided by the number of lattice
sites,$N_{s}$ ) as a function of doping.}
\end{figure}

\section{The projected spin exciton wave function}
In the RPA theory, the resonance mode is interpreted as a spin
exciton: a spin-one particle-hole bound state below the
superconducting gap induced by the residual interaction between the
Bogliubov quasiparticles. The wave function for the spin exciton can
be generally written as
\begin{eqnarray}
|\tilde{\Psi}_{Q}\rangle=\sum_{\mathrm{k}}\phi_{\mathrm{k}}\gamma_{\mathrm{k}\uparrow}^{\dagger}\gamma_{Q-\mathrm{k}\uparrow}^{\dagger}|d-\mathrm{BCS}\rangle,
\end{eqnarray}
in which $\phi_{\mathrm{k}}$ describe the relative motion of the two
quasiparticles $\gamma_{\mathrm{k}\uparrow}^{\dagger}$ and
$\gamma_{Q-\mathrm{k}\uparrow}^{\dagger}$ within the bound state.
$|d-\mathrm{BCS}\rangle$ denotes the BCS mean field ground state and
$\gamma_{\mathrm{k}\uparrow}^{\dagger}$ denotes the creation
operator for the Bogliubov quasiparticles.

As we have shown, the mean field state fails to satisfy the local
spin sum rule. For this reason, we project the spin exciton wave
function Eq.(11) into the subspace of no double occupancy to
construct a variational wave function for the resonance mode, with
$\phi_{\mathrm{k}}$ as the variational parameter to be determined by
optimization of energy.
\begin{eqnarray}
|\Psi_{Q}\rangle=P_{N}P_{G}|\tilde{\Psi}_{Q}\rangle=\sum_{\mathrm{k}}\phi_{\mathrm{k}}|\mathrm{k},Q\rangle,
\end{eqnarray}
in which
\begin{eqnarray}
|\mathrm{k},Q\rangle
=P_{N}P_{G}\gamma_{\mathrm{k}\uparrow}^{\dagger}\gamma_{Q-\mathrm{k}\uparrow}^{\dagger}|d-\mathrm{BCS}\rangle.
\end{eqnarray}
We note that the wave function of the single mode approximation can
also be cast into the form of Eq.(12), \textit{i.e.},
\begin{eqnarray}
|\Psi_{Q}^{0}\rangle&=&
P_{N} P_{G}S_{Q}^{+}|d-\mathrm{BCS} \rangle\nonumber\\
&=&\sum_{\mathrm{k}}P_{N}
P_{G}c_{\mathrm{k}+Q,\uparrow}^{\dagger}c_{\mathrm{k},\downarrow}|d-\mathrm{BCS}
\rangle
\nonumber\\
&=&\sum_{\mathrm{k}}\phi_{\mathrm{k}}^{0}|\mathrm{k},Q\rangle,
\end{eqnarray}
in which $\phi_{\mathrm{k}}^{0}=u_{\mathrm{k}+Q}v_{k}$. Here we have
used the fact that $S_{Q}^{+}$ commute with both $P_{N}$ and
$P_{G}$.

The variational parameters $\phi_{\mathrm{k}}$ can be determined by
minimizing the energy of $|\Psi_{Q}\rangle$ with respect to the
$t-J$ Hamiltonian. The variational energy is given by
\begin{eqnarray}
E_{Q}=\frac{\langle\Psi_{Q}|H_{t-J}|\Psi_{Q}\rangle}{\langle\Psi_{Q}|\Psi_{Q}\rangle}
=\frac{\sum_{\mathrm{k,k'}}\phi_{\mathrm{k}}^{*}H_{\mathrm{k,k'}}\phi_{\mathrm{k'}}}
{\sum_{\mathrm{k,k'}}\phi_{\mathrm{k}}^{*}O_{\mathrm{k,k'}}\phi_{\mathrm{k'}}},
\end{eqnarray}
in which
\begin{eqnarray}
H_{\mathrm{k,k'}}=\langle \mathrm{k},Q|H_{t-J}|\mathrm{k}',Q\rangle
\end{eqnarray}
is the matrix element of the $t-J$ Hamiltonian in the basis
$|\mathrm{k},Q\rangle$ and
\begin{eqnarray}
O_{\mathrm{k,k'}}=\langle \mathrm{k},Q|\mathrm{k}',Q\rangle
\end{eqnarray}
denotes the overlap integral between these non-orthogonal basis
function(note that
$\gamma_{\mathrm{k}\uparrow}^{\dagger}\gamma_{Q-\mathrm{k}\uparrow}^{\dagger}|d-\mathrm{BCS}\rangle$
form a orthogonal basis set before the Gutzwiller projection).

The problem of minimizing $E_{Q}$ respect to the set of variational
parameters $\phi_{\mathrm{k}}$ now reduces to solving the following
generalized eigenvalue problem
\begin{eqnarray}
\sum_{\mathrm{k'}}H_{\mathrm{k,k'}}\phi_{\mathrm{k'}}=\lambda
\sum_{\mathrm{k'}}O_{\mathrm{k,k'}}\phi_{\mathrm{k'}}.
\end{eqnarray}
It is easily seen that the optimized energy $E_{Q}$ is given by the
lowest eigenvalue $\lambda$ of the above generalized eigenvalue
problem.

The above optimization procedure can also be interpreted as
re-diagonalizing the $t-J$ Hamiltonian in the subspace spanned by
the set of non-orthogonal basis function $|\mathrm{k},Q\rangle$.
With this understanding in mind, we can even construct variationally
the full spin fluctuation spectrum as follows
\begin{eqnarray}
S(Q,\omega)=M_{Q}^{2}\sum_{n,\mathrm{k,k'}}|\phi_{\mathrm{k}}^{n*}O_{\mathrm{k,k'}}\phi_{\mathrm{k'}}^{0}|^{2}\delta(\omega-(E_{n}-E_{g})).
\end{eqnarray}
in which $\phi_{\mathrm{k}}^{n}$ denotes the n-th eigenvector of the
generalized eigenvalue problem with the eigenvalue $E_{n}$, $E_{g}$
denotes the variational ground state energy(here we assume that
$\phi_{k}$ and $\phi_{k}^{0}$ are so chosen that both
$|\Psi_{Q}\rangle$ and $|\Psi_{Q}^{0}\rangle$ are normalized). As
$\phi_{\mathrm{k}}^{n}$ forms a orthonormal basis with respect to
the overlap matrix $O_{\mathrm{k,k'}}$, \textit{i.e.},
$\sum_{\mathrm{k,k'}}\phi_{\mathrm{k}}^{n*}O_{\mathrm{k,k'}}\phi_{\mathrm{k'}}^{m}=\delta_{n,m}$,
we have
\begin{eqnarray}
\int d\omega
S(Q,\omega)&=&M_{Q}^{2}\sum_{n}|\sum_{\mathrm{k,k'}}\phi_{\mathrm{k}}^{n*}O_{\mathrm{k,k'}}\phi_{\mathrm{k'}}^{0}|^{2}\nonumber\\
&=&
M_{Q}^{2}\sum_{\mathrm{k,k'}}\phi_{\mathrm{k}}^{0*}O_{\mathrm{k,k'}}\phi_{\mathrm{k'}}^{0}=M_{Q}^{2},
\end{eqnarray}
in which we have used the fact that
\begin{eqnarray}
\sum_{n,\mathrm{k}}\phi_{\mathrm{k_{1}}}^{n*}O_{\mathrm{k_{2},k}}\phi_{\mathrm{k}}^{n}=\delta_{\mathrm{k_{1},k_{2}}},
\end{eqnarray}
which can be derived from the orthonormality of the eigenvectors of
the generalized eigenvalue problem Eq.(18),
\begin{eqnarray}
\sum_{\mathrm{k,k'}}\phi_{\mathrm{k}}^{n*}O_{\mathrm{k,k'}}\phi_{\mathrm{k'}}^{m}=\delta_{n,m}.
\end{eqnarray}
Thus our variational construction of the spin fluctuation spectrum
respects the local spin sum rule of the $t-J$ model.

Our variational scheme for the resonance mode has the advantage that
it involves no tunable parameter: the parameters in
$|\Psi_{g}\rangle$ is determined by optimizing the ground state
energy and the spin exciton wave function $\phi_{\mathrm{k}}$ is
determined by solving the generalized eigenvalue problem. The RPA
correction is thus automatically taken into account in our
formalism.

From the above discussion, we know the single mode approximation can
be taken as a special case of the projected spin exciton wave
function(with
$\phi_{\mathrm{k}}=\phi_{\mathrm{k}}^{0}=u_{\mathrm{k}}v_{Q-\mathrm{k}}$
). Thus by construction, the resonance energy calculated from
Eq.(18) should be lower than that calculated from single mode
approximation. In fact, the mode energy calculated from the single
mode approximation gives the center of gravity of the spin
fluctuation spectrum, while the resonance mode lies at the bottom of
the spectrum. As the single mode approximation already reproduces
the correct trend for the mode energy as a function of doping in the
underdoped regime, we can even expect our variational scheme to
produce quantitatively reasonable result.

\section{The re-weighting technique}
To calculate the energy of the resonance mode, we should first
evaluate the matrix element of the Hamiltonian in the set of
strongly correlated basis functions $|\mathrm{k},Q\rangle$,
$H_{\mathrm{k,k'}}$ and the overlap matrix element
$O_{\mathrm{k,k'}}$. This can be done in principle by the
variational Monte Carlo(VMC) method. For example, to evaluate
$\frac{H_{\mathrm{k,k'}}}{O_{\mathrm{k,k}}}=\frac{\langle
\mathrm{k},Q|H_{t-J}|\mathrm{k'},Q\rangle}{\langle
\mathrm{k},Q|\mathrm{k},Q\rangle}$, we first expand
$|\mathrm{k},Q\rangle$ in an orthogonal basis $|R_{i}\rangle$,
\textit{i.e.},
\begin{eqnarray}
|\mathrm{k},Q\rangle=\sum_{R_{i}}\psi_{\mathrm{k}}(R_{i})|R_{i}\rangle,
\end{eqnarray}
where $\psi(R_{i})$ is the wave function of $|\mathrm{k},Q\rangle$
in this basis. Then we have
\begin{eqnarray}
\frac{H_{\mathrm{k,k'}}}{O_{\mathrm{k,k}}}
=\frac{\sum_{R_{i}}|\psi_{\mathrm{k}}(R_{i})|^{2}
\frac{H\psi_{\mathrm{k'}}(R_{i})}{\psi_{\mathrm{k}}(R_{i})}}{\sum_{R_{i}}|\psi_{\mathrm{k}}(R_{i})|^{2}},
\end{eqnarray}
in which
\begin{eqnarray}
H\psi_{\mathrm{k'}}(R_{i})=\sum_{R_{i'}}\langle
R_{i}|H|R_{i'}\rangle
\frac{\psi_{\mathrm{k'}}(R_{i'})}{\psi_{\mathrm{k}}(R_{i})}.
\end{eqnarray}
Then we sample the basis space $|R_{i}\rangle$ with the weight
$|\psi_{\mathrm{k}}(R_{i})|^{2}$ and do the sum with the standard
VMC technique\cite{Gros}.

The above procedure, though straightforward, is very inefficient. In
our problem, there are $N_{s}^{2}$ Hamiltonian matrix elements
$H_{\mathrm{k,k'}}$ and $N_{s}^{2}$ overlap matrix elements
$O_{\mathrm{k,k'}}$ to be evaluated. For lattice of reasonable size,
say, $14 \times 14$, the number of the matrix elements to be
evaluated would exceed 40000 even take into account the Hermitian
property of $H_{\mathrm{k,k'}}$ and $O_{\mathrm{k,k'}}$. This is
very time consuming. At the same time, the naive approach has the
drawback that it involves large statistical error in the simulation.
This can be seen as follows,
\begin{eqnarray}
\frac{O_{\mathrm{k,k'}}}{O_{\mathrm{k,k}}}
=\frac{\sum_{R_{i}}|\psi_{\mathrm{k}}(R_{i})|^{2}
\frac{\psi_{\mathrm{k'}}(R_{i})}{\psi_{\mathrm{k}}(R_{i})}}{\sum_{R_{i}}|\psi_{\mathrm{k}}(R_{i})|^{2}}.
\end{eqnarray}
Thus, when the node of $\psi_{\mathrm{k}}(R_{i})$ and
$\psi_{\mathrm{k'}}(R_{i})$ do not coincide with each other, we will
run into trouble when we sample $\psi_{\mathrm{k}}(R_{i})$ around
its node as the fluctuation of
$\frac{\psi_{\mathrm{k'}}(R_{i})}{\psi_{\mathrm{k}}(R_{i})}$ becomes
large. A way to reduce the statistical error caused by the
fluctuation of
$\frac{\psi_{\mathrm{k'}}(R_{i})}{\psi_{\mathrm{k}}(R_{i})}$ is to
sample the combined weight
$\mathcal{W}(R_{i})=|\psi_{\mathrm{k}}(R_{i})|^{2}+|\psi_{\mathrm{k'}}(R_{i})|^{2}$
rather than $|\psi_{\mathrm{k}}(R_{i})|^{2}$,
\begin{eqnarray}
\frac{O_{\mathrm{k,k'}}}{O_{\mathrm{k,k}}}
=\frac{O_{\mathrm{k,k'}}}{O_{\mathrm{k,k}}+O_{\mathrm{k',k'}}}\ \ /
\ \ \frac{O_{\mathrm{k,k}}}{O_{\mathrm{k,k}}+O_{\mathrm{k',k'}}}.
\end{eqnarray}
Now the calculation of
$\frac{O_{\mathrm{k,k'}}}{O_{\mathrm{k,k}}+O_{\mathrm{k',k'}}}$ can
be done as
\begin{eqnarray}
\frac{O_{\mathrm{k,k'}}}{O_{\mathrm{k,k}}+O_{\mathrm{k',k'}}}=
\frac{\sum_{R_{i}}\mathcal{W}(R_{i})\frac{\psi_{k}^{*}(R_{i})\psi_{k}(R_{i})}{\mathcal{W}(R_{i})}}
{\sum_{R_{i}}\mathcal{W}(R_{i})}.
\end{eqnarray}
The combined weight samples symmetrically between
$\psi_{\mathrm{k}}(R_{i})$ and $\psi_{\mathrm{k'}}(R_{i})$ and
avoids the fluctuation caused by their uncommon nodes. Thus the
statistical error is much reduced.

The above technique can be easily generalized to calculate all the
$N_{s}^{2}$ overlap matrix elements $O_{\mathrm{k,k'}}$. Here we
sample the combined weight of all the $N_{s}$
$|\psi_{\mathrm{k}}(R_{i})|^{2}$:
$\mathcal{W}(R_{i})=\sum_{\mathrm{k}}|\psi_{\mathrm{k}}(R_{i})|^{2}$.
The calculation is done as follows:
\begin{eqnarray}
\frac{O_{\mathrm{k_{1},k_{2}}}}{\sum_{\mathrm{k}}O_{\mathrm{k,k}}}
=\frac{\sum_{R_{i}}\mathcal{W}(R_{i})\frac{\psi_{\mathrm{k_{1}}}^{*}(R_{i})\psi_{\mathrm{k_{2}}}(R_{i})}{\mathcal{W}(R_{i})}}{\sum_{R_{i}}\mathcal{W}(R_{i})}.
\end{eqnarray}
To sample $\mathcal{W}(R_{i})$, we note that
\begin{eqnarray}
\mathcal{W}(R_{i})=|\psi_{\mathrm{k_{0}}}(R_{i})|^{2}
\sum_{\mathrm{k}}|\frac{\psi_{\mathrm{k}}(R_{i})}{\psi_{\mathrm{k_{0}}}(R_{i})}|^{2}
\end{eqnarray}
and
\begin{eqnarray}
\frac{\psi_{\mathrm{k_{1}}}^{*}(R_{i})\psi_{\mathrm{k_{2}}}(R_{i})}{\mathcal{W}(R_{i})}
=
\frac{(\frac{\psi_{\mathrm{k_{1}}}(R_{i})}{\psi_{\mathrm{k_{0}}}(R_{i})})^{*}
\times
\frac{\psi_{\mathrm{k_{2}}}(R_{i})}{\psi_{\mathrm{k_{0}}}(R_{i})}}
{\sum_{\mathrm{k}}|\frac{\psi_{\mathrm{k}}(R_{i})}{\psi_{\mathrm{k_{0}}}(R_{i})}|^{2}},
\end{eqnarray}
in which $\psi_{\mathrm{k_{0}}}$ is one basis function arbitrarily
chosen from the $N_{s}$ basis functions. From this transformation,
we see all we need to calculate in order to evaluate
$\frac{O_{\mathrm{k_{1},k_{2}}}}{\sum_{\mathrm{k}}O_{\mathrm{k,k}}}$
is the $N_{s}$ ratio between basis functions
$\frac{\psi_{\mathrm{k}}(R_{i})}{\psi_{\mathrm{k_{0}}}(R_{i})}$. As
the different basis functions are all Slater determinant differing
with each other by at most in a pair of quasiparticle excitations,
such ratio is easy to calculate using the inverse updating technique
for Fermion determinant. More importantly, the calculation of all
the $N_{s}^{2}$ overlap matrix elements can be done in a single
Monte Carlo simulation: the algorithm is highly parallelized.

The calculation of the Hamiltonian matrix elements can be done
similarly. We have
\begin{eqnarray}
\frac{H_{\mathrm{k_{1},k_{2}}}}{\sum_{\mathrm{k}}O_{\mathrm{k,k}}}
=\frac{\sum_{R_{i}}\mathcal{W}(R_{i})\frac{\psi_{\mathrm{k_{1}}}^{*}(R_{i})\times
\
 H\psi_{\mathrm{k_{2}}}(R_{i})}{\mathcal{W}(R_{i})}}{\sum_{R_{i}}\mathcal{W}(R_{i})},
\end{eqnarray}
where $H\psi_{\mathrm{k_{2}}}(R_{i})=\sum_{R_{j}}\langle
R_{i}|H_{t-J}|R_{j}\rangle\psi_{\mathrm{k_{2}}}(R_{j})$. Following
the same reasoning, we arrive at
\begin{eqnarray}
\frac{\psi_{\mathrm{k_{1}}}^{*}(R_{i})\times
 \ H\psi_{\mathrm{k_{2}}}(R_{i})}{\mathcal{W}(R_{i})} =
\frac{(\frac{\psi_{\mathrm{k_{1}}}(R_{i})}{\psi_{\mathrm{k_{0}}}(R_{i})})^{*}
\times
\frac{H\psi_{\mathrm{k_{2}}}(R_{i})}{\psi_{\mathrm{k_{0}}}(R_{i})}}
{\sum_{\mathrm{k}}|\frac{\psi_{\mathrm{k}}(R_{i})}{\psi_{\mathrm{k_{0}}}(R_{i})}|^{2}}.
\end{eqnarray}
Thus the calculation of the Hamiltonian matrix elements involves the
evaluation of the ratio
$\frac{H\psi_{\mathrm{k_{2}}}(R_{i})}{\psi_{\mathrm{k_{0}}}(R_{i})}$.
The calculation of this ratio, though numerically more demanding, is
still highly parallelized.

Thus the re-weighting technique developed here not only reduce
considerably the statistical error involved in the Monte Carlo
simulation, but also highly parallelize the calculation of the
overlap and Hamiltonian matrix elements, reducing their calculation
from the order of $N_{s}^{2}$ to a single Monte Carlo simulation.

\section{Numerical results}
To calculate the energy of the neutron resonance mode, we first
optimize the variational parameters $t'_{v}$, $\mu_{v}$,
$\Delta_{v}$ as a function of doping for the ground state. The
calculation is done on a $14 \times 14$ lattice with
periodic-antiperiodic boundary condition. We choose
$\frac{J}{t}=\frac{1}{3}$ and $\frac{t'}{t}=-0.25$ in the $t-J$
model to describe a hole doped cuprate. The results of the optimized
variational parameters as a function of doping are shown in Fig.2
and Fig.3.

\begin{figure}[h!]
\includegraphics[width=8cm,angle=0]{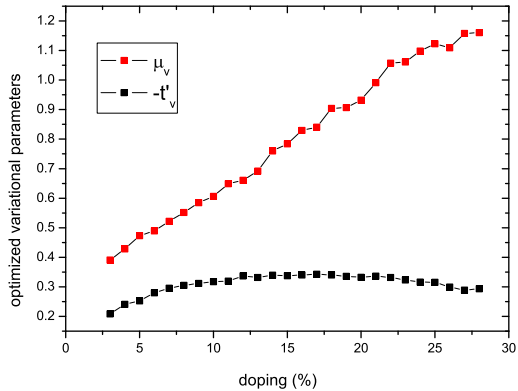}
\caption{The optimized variational parameters as a function of
doping.}
\end{figure}

\begin{figure}[h!]
\includegraphics[width=8cm,angle=0]{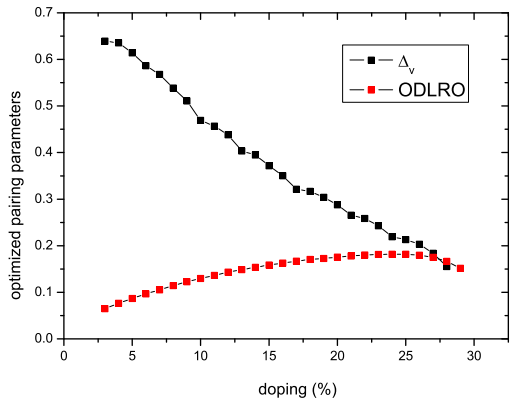}
\caption{The optimized pairing parameters as a function of doping
and the off diagonal long range order calculated from the
variational ground state.}
\end{figure}

\begin{figure}[h!]
\includegraphics[width=6cm,angle=0]{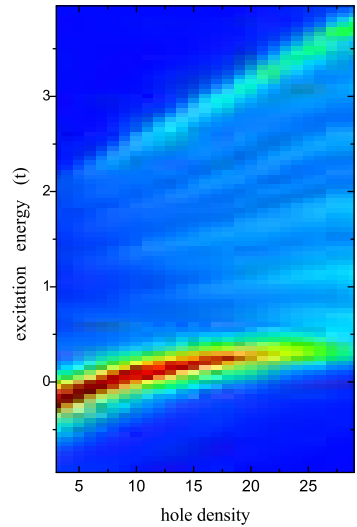}
\caption{The spin fluctuation spectrum at $q=(\pi,\pi)$ as a
function of excitation energy and doping determined from the
variational calculation. The calculation is done on a $14 \times 14$
lattice. The delta function peaks of the spectrum are broadened into
Lorentzian peaks with a width of $0.1t$. }
\end{figure}

We note the superconducting region determined by the variational
approach for the $t-J$ model is considerably larger than that
observed in experiments. In Fig.3, we plot the off diagonal long
range order(ODLRO) calculated from the optimized variational ground
state. The ODLRO is defined as
\begin{eqnarray}
\Delta=\sqrt{\frac{1}{N_{s}}\sum_{i}\langle\hat{\Delta}_{i}\hat{\Delta}^{\dagger}_{i+R_{M}},
\rangle}
\end{eqnarray}
in which
$\hat{\Delta}_{i}=(c_{i+x,\downarrow}c_{i,\uparrow}-c_{i+x,\uparrow}c_{i,\downarrow})-(c_{i+y,\downarrow}c_{i,\uparrow}-c_{i+y,\uparrow}c_{i,\downarrow})$
is the Cooper pair annihilation operator at site $i$, $R_{M}$ is the
largest distance on the finite lattice. We find the ODLRO reaches
its maximal around $x\approx 26\%$. We will take this doping
concentration as an estimate of the location of the optimal doping
in the following discussion.

 After obtaining the variational parameters of the
ground state we are now ready to calculate the Hamiltonian matrix
elements $H_{\mathrm{k,k'}}$ and the overlap matrix elements
$O_{\mathrm{k,k'}}$. This is the most heavy part of our numerical
calculation. In our calculation, we have sampled more than two and a
half million configurations with the weight $\mathcal{W}(R_{i})$.
The accept ratio is tuned to be $\frac{1}{2}$.  The statistical
error is found to be smaller than the fluctuation caused by the
finite size effect in our calculation.

When we get the matrix elements $H_{\mathrm{k,k'}}$ and
$O_{\mathrm{k,k'}}$, we solve the generalized eigenvalue problem
Eq.(18) and calculate the physical quantities that we are interested
in. First we show the spin fluctuation spectrum calculated in this
variational approach in Fig.4. We see the spectrum consists of both
a coherent peak and a continuum of incoherent spin fluctuation. The
coherent peak at the bottom of the spectrum is nothing but the
neutron resonance mode in our variational description.

\begin{figure}[h!]
\includegraphics[width=8cm,angle=0]{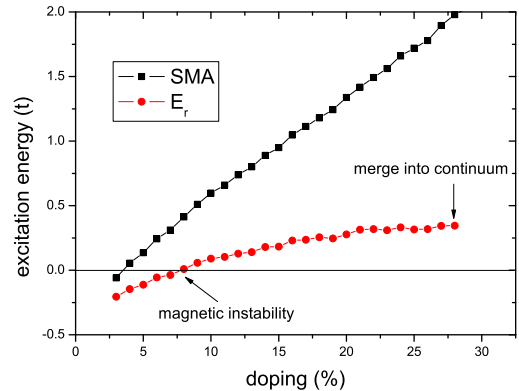}
\caption{The mode energy determined by our variational approach as a
function of doping as compared with the result of single mode
approximation(SMA).}
\end{figure}

Fig.5 shows the mode energy as a function of doping. We find the
mode energy becomes negative below a critical doping around
$x=7.5\%$. A negative excitation energy indicates magnetic
instability of the system. The critical doping so determined is
close but sightly lower than that determined by assuming directly a
magnetic order in the variational ground state, which is about
$10\%$\cite{Ogata}. In whole doping range in which the mode has
nonzero spectral weight, the mode energy is a monotonically
increasing function of doping and reaches about $0.3t$ before it
loses its weight and merges into the particle-hole continuum at
about $x=29\%$. If we take $t=0.25eV$ as is usually done in the
literature, we get the maximum of the mode energy to be about $75
meV$, about a factor of 1.8 larger than that observed in optimally
doped $YBa_{2}Cu_{3}O_{6.93}$\cite{Fong}.

\begin{figure}[h!]
\includegraphics[width=8cm,angle=0]{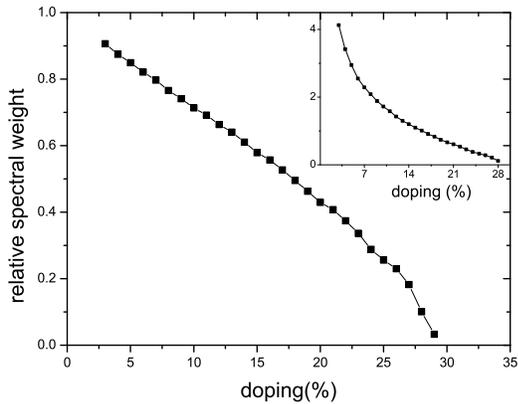}
\caption{The relative spectral weight of the resonance mode as a
function of doping. The inset shows the result of the absolute
spectral weight.}
\end{figure}
In our theory, the mode energy increases monotonically with doping.
This is consistent with experiments in the underdoped regime but may
have inconsistency with experiment in the overdoped regime, where
experiment reported evidence of a weak resonance mode with an energy
sightly lower than that of the optimally doped system\cite{He}. If
we take the doping at which the ODLRO reaches maximum($x=26\%$) as
the optimal doping, then the resonance mode will survive in the
slightly overdoped regime with an energy higher than that of the
optimal doped system. The would imply a breakdown of the linear
scaling between the mode energy and $T_{c}$ in the slightly
overdoped regime.

Fig.6 shows the relative and absolute spectral weight of the
resonance mode calculated from our theory. The relative spectral
weight is defined as the proportion of the mode intensity to the
total spectral weight at $Q=(\pi,\pi)$,
\begin{eqnarray}
W=\frac{|\langle\Psi_{Q}^{0}|\Psi_{Q}\rangle|^{2}}{\langle\Psi_{Q}^{0}|\Psi_{Q}^{0}\rangle}.
\end{eqnarray}
The absolute spectral weight is defined as the product of the
relative spectral weight and the spin structure factor divided by
$N_{s}$. The absolute spectral weight decreases rapidly with doping
as a result of the decrease of both the magnetic structure factor
and the relative spectral weight. The relative spectral weight of
the resonance mode decrease from unity at half filling down to zero
at about $x=29\%$. The unity of the relative spectral weight come
from the long range correlation of spin at half filling and
indicates that the resonance mode can be connected smoothly to the
Goldstone mode in the magnetic ordered state.

\section{Discussion}
In this paper, we proposed a variational theory for the neutron
resonance mode in the cuprate superconductors. Our theory has the
virtue that it involves no free parameter and thus has much larger
predictive power than the phenomenological RPA treatment of the spin
fluctuation. In our theory, the RPA correction is automatically
taken into account by the variational procedure which reduces to
solving a generalized eigenvalue problem. More importantly, our
variational approach builds in the no double occupancy constraint
and thus satisfies the local spin sum rule of the $t-J$ model. This
is of vital importance for a correct description of the spin
dynamics of cuprates. Our approach has the further advantage that it
provides a physical transparent understanding of the resonance mode
as a spin exciton in the physical subspace of no double occupancy.

Our approach can also be taken as diagonalizing the Hamiltonian in a
truncated subspace with the same quantum number as the excitation
discussed and can be used to calculate the full spectrum rather than
the coherent excitation only. It is important to note this truncated
subspace exhaust the spectral weight for the relevant sum rule,
indicating the relevance of the spectrum calculated in this way. An
effort to apply the current approach to calculate the spin
fluctuation spectrum at momentum other than $(\pi,\pi)$, namely the
incommensurate spin fluctuation spectrum, is now under
investigation.

We have also devised a very efficient re-weighting technique to
tackle the numerical problem of simulating the variational wave
function composed of N linearly superposed Slater determinants. The
key for the efficiency of the algorithm is the observation that the
Slater determinants involved in our wave function differ with each
other by at most a pair of quasiparticle excitations. Obviously this
technique can be applied in a much larger literature than simulating
the physics of cuprates.

As we have found in the single mode approximation, the center of
gravity of the spin fluctuation spectrum increases monotonically
with doping as a result of the rapid decrease of the spin structure
factor at $Q=(\pi,\pi)$. We find the resonance mode, which lies at
the bottom of the spin fluctuation spectrum, inherits this monotonic
behavior, probably for the same reason. As we have mentioned above,
this may have potential conflict with the report of weak resonance
mode in the overdoped sample with an energy lower than that of the
optimal doped system\cite{He} and would imply the breakdown of the
$T_{c}-E_{r}$ linear scaling in the overdoped regime.

In our theory, the monotonic increase of the mode energy eventually
cut off at about twice the maximal superconducting gap when the mode
transfers all of its weight into the particle-hole continuum. To
exhibit a non-monotonic behavior before merging into the
particle-hole continuum, it is necessary for the superconducting gap
to decrease faster with doping than that predicted by the present
variational calculation. This is not at all impossible. However,
since there is no generally reason to believe the $T_{c}-E_{r}$
linear scaling to hold in the overdoped regime and the mode weight
become very small in the overdoped regime, we think the mode energy
in the overdoped regime is a problem subjected to fine-tuning.

The mode energy at the optimal doping($x=26\%$) as calculated from
our approach is about $0.3t$ and is a factor of 1.8 larger than that
observed in optimally doped $YBa_{2}Cu_{3}O_{6.93}$. At the same
time, the variational theory predicts a considerably larger value of
optimal doping than observed in experiments($x=16\%$). It is likely
that these two problems and the problem of the $T_{c}-E_{r}$ linear
scaling to hold in the overdoped regime are related with each other.
However, it is not clear to what extend should we attribute these
disagreements with experiment solely to the limitation of the
variational approach we have adopted rather than the intrinsic
properties of the $t-J$ model.

Finally, we discuss the relation between the resonance mode and the
superconductivity. As many other theories of the resonance
mode\cite{Liu,Weng,Lee}, our theory also predicts that the resonance
mode becomes stronger and stronger with decreasing doping and
evolves smoothly into the Goldstone mode of the ordered state at
half filling. On the other hand, in the phenomenological SO(5)
theory of cuprates\cite{ZhangSC}, in which the mode is understood as
a pseudo-Goldstone mode accompanying breaking of an SO(5) symmetry
between the d-wave superconducting order and the antiferromagnetic
order, the mode intensity is predicted to be proportional to the off
diagonal long range order(ODLRO) of the system. Thus the theory
predicts that the mode intensity should decrease when we increase
temperature or decrease doping and disappear out of the
superconducting dome. On the experimental side, the resonance mode
is observed to loss weight with increasing temperature and to
disappear in the normal state at optimal doping. In slightly
underdoped sample, a broadened and weak signal is observed above
$T_{c}$. With further decrease of doping, the normal state signal
becomes stronger and stronger and the enhancement due to
superconductivity becomes less and less
prominent\cite{Fong,Fong1,Dai}.

In our theory, we only consider the zero temperature case. At finite
temperature, both quasiparticle excitations and collective
fluctuation of the superconducting order parameter will be thermally
excited. The latter excitation is believed to be especially
important around the superconducting transition point. We believe
both of these thermal excitations are responsible for the decrease
of the mode intensity with increasing temperature. The doping
dependence of the mode intensity is more subtle. Here the relevant
question is why the superconductivity-related enhancement of the
mode intensity becomes smaller and smaller with decreasing doping.
In the very underdoped regime, the electron correlation(Mott
physics)is greatly enhanced. We believe the Mott physics is at work
in reducing the superconductivity-related enhancement of the mode
intensity in the very underdoped regime. It is interesting to note
that in the RVB picture, the electron correlation effect manifests
itself as quantum fluctuation of the superconducting order
parameter.

In all, apart from some subtle issues mentioned above, our
variational approach capture the gross feature of the neutron
resonance mode and provides the first truly microscopic
understanding of this important phenomena in the high-temperature
superconductors. As a by product of this research, we developed a
very efficient re-weighting technique to simulated wave function
composed of N linearly superposed Slater determinants, an algorithm
whose potential application is obviously far beyond the high-Tc
issue.

T. Li is supported by NSFC Grant No. 10774187, National Basic
Research Program of China No. 2007CB925001 and Beijing Talent
Program. F. Yang is supported by NSFC Grant No. 10704008.

\end{document}